DATA WAREHOUSE BENCHMARKING WITH DWEB

Jérôme Darmont
University of Lyon (ERIC Lyon 2)
5 avenue Pierre Mendès-France
69676 Bron Cedex
France
*jerome.darmont@univ-lyon2.fr*

ABSTRACT

Performance evaluation is a key issue for designers and users of Database Management Systems (DBMSs). Performance is generally assessed with software benchmarks that help, e.g., test architectural choices, compare different technologies or tune a system. In the particular context of data warehousing and On-Line Analytical Processing (OLAP), although the Transaction Processing Performance Council (TPC) aims at issuing standard decision-support benchmarks, few benchmarks do actually exist. We present in this chapter the Data Warehouse Engineering Benchmark (DWEB), which allows generating various ad-hoc synthetic data warehouses and workloads. DWEB is fully parameterized to fulfill various data warehouse design needs. However, two levels of parameterization keep it relatively easy to tune. We also expand on our previous work on DWEB by presenting its new Extract, Transform, and Load (ETL) feature as well as its new execution protocol. A Java implementation of DWEB is freely available on-line, which can be interfaced with most existing relational DMBSs. To the best of our knowledge, DWEB is the only easily available, up-to-date benchmark for data warehouses.

INTRODUCTION

Performance evaluation is a key issue for both designers and users of Database Management Systems (DBMSs). It helps designers select among alternate software architectures, performance optimization strategies, or validate or refute hypotheses regarding the actual behavior of a system. Thus, performance evaluation is an essential component in the development process of efficient and well-designed database systems. Users may also employ performance evaluation, either to compare the efficiency of different technologies before selecting one, or to tune a system. In many fields including databases, performance is generally assessed with the help of software benchmarks. The main components in a benchmark are its database model and workload model (set of operations to execute on the database).

Evaluating data warehousing and decision-support technologies is a particularly intricate task. Though pertinent, general advice is available, notably on-line (Pendse, 2003; Greenfield, 2004a), more quantitative elements regarding sheer performance, including benchmarks, are few. In the late nineties, the OLAP (On-Line Analytical Process) APB-1 benchmark has been very popular. Henceforth, the Transaction Processing Performance Council (TPC) (1), a non-profit organization, defines standard benchmarks (including decision-support benchmarks) and publishes objective and verifiable performance evaluations to the industry.

Our own motivation for data warehouse benchmarking was initially to test the efficiency of performance optimization techniques (such as automatic index and materialized view selection techniques) we have been developing for several years. None of the existing data warehouse benchmarks suited our needs. APB-1's schema is fixed, while we needed to test our performance optimization techniques on various data warehouse configurations. Furthermore, it is no longer supported and somewhat difficult to find. The TPC currently supports the TPC-H decision-support benchmark (TPC, 2006). However, its database schema is inherited from the older and obsolete benchmark TPC-D (TPC, 1998), which is not a dimensional schema such as the typical star schema and its derivatives that are used in data warehouses (Inmon, 2002; Kimball & Ross, 2002). Furthermore, TPC-H's workload, though decision-oriented, does not include explicit OLAP queries either. This benchmark is implicitly considered obsolete by the TPC that has issued some draft specifications for its successor: TPC-DS (TPC, 2007). However, TPC-DS, which is very complex, especially at the ETL (Extract, Transform, and Load) and workload levels, has been under development since 2002 and is not completed yet.

Furthermore, although the TPC decision-support benchmarks are scalable according to Gray's (1993) definition, their schema is also fixed. For instance, TPC-DS' constellation schema cannot easily be simplified into a simple star schema. It must be used "as is". Different ad-hoc configurations are not possible. Furthermore, there is only one parameter to define the database, the Scale Factor ($SF$), which sets up its size (from 1 to 100,000 GB). Users cannot control the size of dimensions and fact tables separately, for instance. Finally, users have no control on workload definition. The number of generated queries directly depends on $SF$.

Eventually, in a context where data warehouse architectures and decision-support workloads depend a lot on application domain, it is very important that designers who wish to evaluate the impact of architectural choices or optimization techniques on global performance can choose and/or compare among several configurations. The TPC benchmarks, which aim at standardized results and propose only one configuration of warehouse schema, are ill-adapted to this purpose. TPC-DS is indeed able to evaluate the performance of optimization techniques, but it cannot test their impact on various choices of data warehouse architectures. Generating particular data warehouse configurations (e.g., large-volume dimensions) or ad-hoc query workloads is not possible either, whereas it could be an interesting feature for a data warehouse benchmark.

For all these reasons, we decided to design a full data warehouse benchmark that would be able to model various configurations of database and workload: DWEB, the Data Warehouse Engineering Benchmark (Darmont *et al.*, 2005; Darmont *et al.*, 2007). In this context (variable architecture, variable size), using a real-life benchmark is not an option. Hence, DWEB helps generate ad-hoc synthetic data warehouses (modeled as star, snowflake, or constellation schemas) and workloads, mainly for engineering needs. DWEB may thus be viewed more as a benchmark generator than an actual, single benchmark.

This chapter presents the full specifications of DWEB's database and workload models, and expands our previous work with a new ETL process and a new execution protocol that have recently been included into DWEB. All models, parameters and pseudo-code algorithms are provided. The remainder of this chapter is organized as follows. We first present the state of the art decision-support benchmarks, with a particular focus on the current and future standards TPC-H and TPC-DS. Then, we detail DWEB's specifications: database model, workload model, ETL process and execution protocol. We present a short tutorial to illustrate DWEB's usage, and finally conclude this chapter and provide future research directions.

## STATE OF THE ART DECISION-SUPPORT BENCHMARKS

To the best of our knowledge, relatively few decision-support benchmarks have been designed out of the TPC. Some do exist, but their specification is sometimes not fully published (Demarest, 1995). The most notable is presumably the OLAP APB-1 benchmark, which was issued in 1998 by the OLAP council, a now inactive organization founded by four OLAP vendors. APB-1 has been quite extensively used in the late nineties. Its data warehouse schema is architectured around four dimensions: *Customer*, *Product*, *Channel* and *Time*. Its work-load of ten queries is aimed at sale forecasting. APB-1 is quite simple and proved limited, since it is not "differentiated to reflect the hurdles that are specific to different industries and functions" (Thomsen, 1998). Finally, some OLAP datasets are also available on-line (2), but they do not qualify as benchmarks, being only raw databases (chiefly, no workload is provided).

In the remainder of this section, we focus more particularly on the TPC benchmarks. The TPC-D benchmark (Ballinger, 1993; Bhashyam, 1996; TPC, 1998) appeared in the mid-nineties, and forms the base of TPC-H and TPC-R that have replaced it (Poess & Floyd, 2000). TPC-H and TPC-R are actually identical, only their usage varies. TPC-H (TPC, 2006) is for ad-hoc querying (queries are not known in advance and optimizations are forbidden), while TPC-R (TPC, 2003) is for reporting (queries are known in advance and optimizations are allowed). TPC-H is currently the only decision-support benchmark supported by the TPC. Its designated successor, TPC-DS (Poess *et al.*, 2002; TPC, 2007), is indeed still currently under development and only available as a draft.

TPC-H (Note: This is actually a subheading, but the acronym is wholly in uppercase)

TPC-H exploits the same relational database schema as TPC-D: a classical *product-order-supplier* model (represented as a UML class diagram in Figure 1); and the workload from TPC-D supplemented with five new queries. This workload is constituted of twenty-two SQL-92 parameterized, decision-oriented queries labeled Q1 to Q22; and two refresh functions RF1 and RF2 that essentially insert and delete tuples in the ORDER and LINEITEM tables.

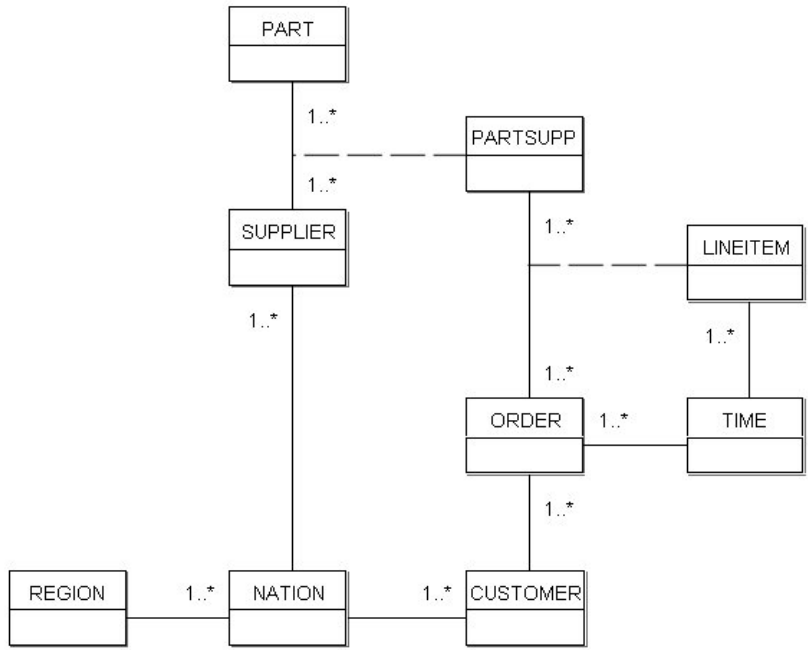

Figure 1: TPC-D, TPC-H, and TPC-R database schema

The query parameters are substituted with the help of a random function following a uniform distribution. Finally, the protocol for running TPC-H includes a load test and a performance test (executed twice), which is further subdivided into a power test and a throughput test. Three primary metrics describe the results in terms of power, throughput, and a composition of the two. Power and throughput are respectively the geometric and arithmetic average of database size divided by execution time.

TPC-DS (Note: This is actually a subheading, but the acronym is wholly in uppercase)

TPC-DS more clearly models a data warehouse than TPC-H. TPC-DS' database schema, whose fact tables are represented in Figure 2, models the decision-support functions of a retail product supplier as several snowflake schemas. Catalog and web sales and returns are interrelated, while store management is independent. This model also includes fifteen dimensions that are shared by the fact tables. Thus, the whole model is a constellation schema.

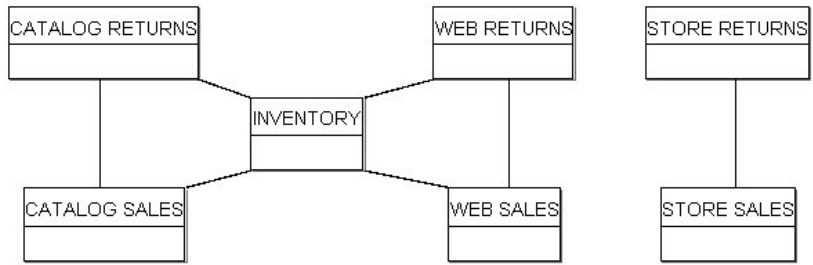

Figure 2: TPC-DS data warehouse schema

TPC-DS' workload is made of four classes of queries: reporting queries, ad-hoc decision-support queries, interactive OLAP queries, and data extraction queries. A set of about five hundred queries is generated from query templates written in SQL-99 (with OLAP extensions). Substitutions on the templates are operated using non-uniform random distributions. The data warehouse maintenance process includes a full ETL process and a specific treatment of dimensions. For instance, historical dimensions preserve history as new dimension entries are added, while non-historical dimensions do not store aged data any more. Finally, the execution model of TPC-DS consists of four steps: a load test, a query run, a data maintenance run, and another query run. A single throughput metric is proposed, which takes the query and maintenance runs into account.

DWEB SPECIFICATIONS

We present in this section the fullest specifications of DWEB as of today. The main components in a benchmark are its database and workload models, but we also detail DWEB's new ETL capability and new execution protocol, which were previously assumed to be similar to TPC-DS's.

Database Model

*Schema* (Note: This is a sub-subheading)

Our design objective for DWEB is to be able to model the different kinds of data warehouse architectures that are popular within a ROLAP (Relational OLAP) environment: classical star schemas, snowflake schemas with hierarchical dimensions, and constellation schemas with multiple fact tables and shared dimensions. To achieve this goal, we propose a data ware-house metamodel (represented as a UML class diagram in Figure 3) that can be instantiated into these different schemas.

We view this metamodel as a middle ground between the multidimensional metamodel from the Common Warehouse Metamodel (CWM) (OMG, 2003; Poole *et al.*, 2003) and the eventual benchmark model. Our metamodel may actually be viewed as an instance of the CWM metamodel, which could be qualified as a meta-metamodel in our context. The upper part of Figure 3 describes a data warehouse (or a datamart, if a datamart is viewed as a small, dedicated data warehouse) as constituted of one or several fact tables that are each described by several dimensions. Each dimension may also describe several fact tables (shared dimensions). Each dimension may be constituted of one or several hierarchies made of different levels. There can be only one level if the dimension is not a hierarchy. Both fact tables and dimension hierarchy levels are relational tables, which are modeled in the lower part of Figure 3. Classically, a table or relation is defined in intention by its attributes and in extension by its tuples or rows. At the intersection of a given attribute and a given tuple lies the value of this attribute in this tuple.

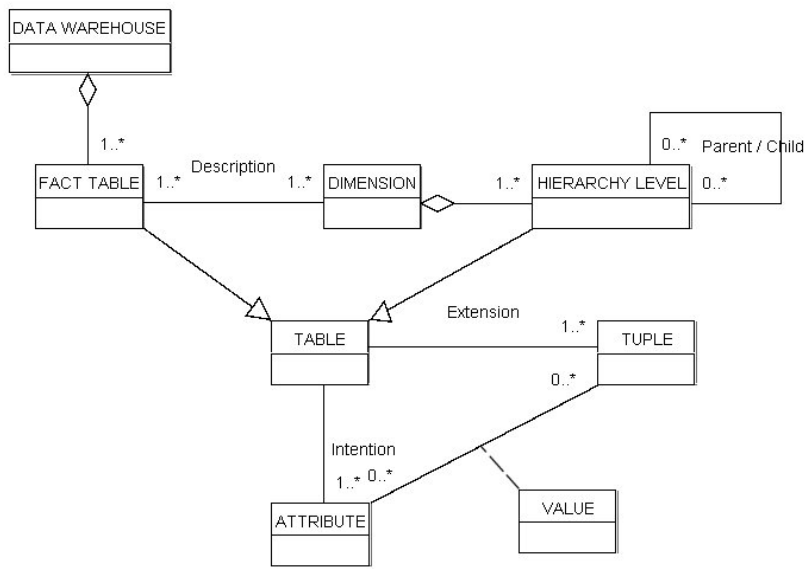

Figure 3: DWEB data warehouse metaschema

Our metamodel is quite simple. It is sufficient to model the data warehouse schemas we aim at (star, snowflake, and constellation schemas), but it is limited and cannot model some particularities that are found in real-life warehouses, such as many-to-many relationships between facts and dimensions, or hierarchy levels shared by several hierarchies. This is currently a deliberate choice, but the metamodel might be extended in the future.

*Parameterization*

DWEB's database parameters help users select the data warehouse architecture they need in a given context. They are aimed at parameterizing metaschema instantiation to produce an actual data warehouse schema. When designing them, we try to meet the four key criteria that make a "good" benchmark, according to Gray (1993):

- *relevance,* the benchmark must answer our engineering needs (cf. Introduction);

- *portability,* the benchmark must be easy to implement on different systems;

- *scalability,* it must be possible to benchmark small and large databases, and to scale up the benchmark;

- *simplicity,* the benchmark must be understandable; otherwise it will not be credible nor used.

We further propose to extend Gray's scalability criterion to *adaptability*. A performance evaluation tool must then propose different database or workload configurations that help run tests in various experimental conditions. Such tools might be qualified as benchmark generators, though we term them, possibly abusively, tunable or generic benchmarks. Aiming at adaptability is mechanically detrimental to simplicity. However, this criterion is fundamental and must not be neglected when designing a generic tool. It is thus necessary to find means to ensure good adaptability while not sacrificing simplicity; in short, to find a fair tradeoff between these criteria.

Relevance and adaptability on one hand, and simplicity on the other hand, are clearly two orthogonal goals. Introducing too few parameters reduces the model's expressiveness, while introducing too many parameters makes it difficult to apprehend by potential users. Furthermore, few of these parameters are likely to be used in practice. In parallel, the generation complexity of the instantiated schema must be mastered. To solve this dilemma, we capitalize on the experience of designing the OCB object-oriented database benchmark (Darmont & Schneider, 2000). OCB is generic and able to model all the other existing object-oriented database benchmarks, but it is controlled by too many parameters, few of which are used in practice. Hence, we propose to divide the parameter set into two subsets.

The first subset of so-called low-level parameters allows an advanced user to control everything about data warehouse generation (Table 1). However, the number of low-level parameters can increase dramatically when the schema gets larger. For instance, if there are several fact tables, all their characteristics, including dimensions and their own characteristics, must be defined for each fact table.

| Parameter name | Meaning |
|---|---|
| NB_FT | Number of fact tables |
| NB_DIM(f) | Number of dimensions describing fact table #f |
| TOT_NB_DIM | Total number of dimensions |
| NB_MEAS(f) | Number of measures in fact table #f |
| DENSITY(f) | Density rate in fact table #f |
| NB_LEVELS(d) | Number of hierarchy levels in dimension #d |
| NB_ATT(d,h) | Number of attributes in hierarchy level #h of dimension #d |
| HHLEVEL_SIZE(d) | Cardinality of the highest hierarchy level of dimension #d |
| DIM_SFACTOR(d) | Size scale factor in the hierarchy levels of dimension #d |

Table 1: DWEB warehouse low-level parameters

Thus, we designed a layer above with much fewer parameters that may be easily understood and set up (Table 2). More precisely, these high-level parameters are average values for the low-level parameters. At database generation time, the high-level parameters are exploited by random functions (following a Gaussian distribution) to automatically set up the low-level parameters. Finally, unlike the number of low-level parameters, the number of high-level parameters always remains constant and reasonable (less than ten parameters).

| Parameter name | Meaning | Default value |
|---|---|---|
| AVG_NB_FT | Average number of fact tables | 1 |
| AVG_NB_DIM | Average number of dimensions per fact table | 5 |
| AVG_TOT_NB_DIM | Average total number of dimensions | 5 |
| AVG_NB_MEAS | Average number of measures in fact tables | 5 |
| AVG_DENSITY | Average density rate in fact tables | 0.6 |
| AVG_NB_LEVELS | Average number of hierarchy levels in dimensions | 3 |
| AVG_NB_ATT | Average number of attributes in hierarchy levels | 5 |
| AVG_HHLEVEL_SIZE | Average cardinality of the highest hierarchy levels | 10 |
| DIM_SFACTOR | Average size scale factor within hierarchy levels | 10 |

Table 2: DWEB warehouse high-level parameters

Users may choose to set up either the full set of low-level parameters, or only the high-level parameters, for which we propose default values that correspond to a snowflake schema. Note that these parameters control both schema and data generation.

<u>Remarks:</u>

- Since shared dimensions are possible, $TOT\_NB\_DIM \leq \sum_{i=1}^{NB\_FT} NB\_DIM(i)$.

- The cardinal of a fact table is usually lower or equal to the product of its dimensions' cardinals. This is why we introduce the notion of density. A density rate of one indicates that all the possible combinations of the dimension primary keys are present in the fact table, which is very rare in real-life data warehouses. When density rate decreases, we progressively eliminate some of these combinations (cf. Workload Generation).

- This parameter helps control the size of the fact table, independently of the size of its dimensions, which are defined by the *HHLEVEL_SIZE* and *DIM_SFACTOR* parameters (see below).

- Within a dimension, a given hierarchy level has normally a greater cardinality than the next level. For example, in a *town-region-country* hierarchy, the number of towns must be greater than the number of regions, which must be in turn greater than the number of countries. Furthermore, there is often a significant scale factor between these cardinalities (e.g., one thousand towns, one hundred regions, ten countries). Hence, we model the cardinality of hierarchy levels by assigning a "starting" cardinality to the highest level in the hierarchy (*HHLEVEL_SIZE*), and then by multiplying it by a predefined scale factor (*DIM_SFACTOR*) for each lower-level hierarchy.

- The global size of the data warehouse is assessed at generation time so that the user retains full control over it.

*Generation Algorithm*

The instantiation of the DWEB metaschema into an actual benchmark schema is done in two steps:
1. build the dimensions;
2. build the fact tables.

The pseudo-code for these two steps is provided in Figures 4 and 5, respectively. Each of these steps is further subdivided, for each dimension or each fact table, into generating its intention and extension. In addition, dimension hierarchies must be managed. Note that they are generated starting from the highest level of hierarchy. For instance, for our *town-region-country* sample hierarchy, we build the country level first, then the region level, and eventual-ly the town level. Hence, tuples from a given hierarchy level can refer to tuples from the next level (that are already created) with the help of a foreign key.

```
For i = 1 to TOT_NB_DIM do
    previous_ptr = NIL
    size = HHLEVEL_SIZE(i)
    For j = 1 to NB_LEVELS(i) do
        // Intention
        h1 = New(Hierarchy_level)
        h1.intention = Primary_key()
        For k = 1 to NB_ATT(i,j) do
            h1.intention = h1.intention
                ∪ String_member()
        End for
        // Hierarchy management
        h1.child = previous_ptr
        h1.parent = NIL
        If previous_ptr ≠ NIL then
            previous_ptr.parent = h1
            // Foreign key
            h1.intention = h1.intention
                ∪ previous_ptr.intention.primary_key
        End if
        // Extension
        h1.extension = ∅
        For k = 1 to size do
            new_tuple = Integer_primary_key()
            For l = 1 to NB_ATT(i,j) do
                new_tuple = new_tuple ∪ Random_string()
            End for
            If previous_ptr ≠ NIL then
                new_tuple = new_tuple
                    ∪ Random_key(previous_ptr)
            End if
            h1.extension = h1.extension ∪ new_tuple
        End for
        previous_ptr = h1
        size = size * DIM_SFACTOR(i)
    End for
    dim(i) = h1 // First (lowest) level of the hierarchy
End for
```

Figure 4: DWEB dimensions generation algorithm

```
For i = 1 to TOT_NB_FT do
    // Intention
    ft(i).intention = ∅
    For k = 1 to NB_DIM(i) do
        j = Random_dimension(ft(i))
        ft(i).intention = ft(i).intention
            ∪ ft(i).dim(j).primary key
    End for
    For k to NB_MEAS(i) do
        ft(i).intention = ft(i).intention
            ∪ Float_measure()
    End for
    // Extension
    ft(i).extension = ∅
    For j = 1 to NB_DIM(i) do // Cartesian product
        ft(i).extension = ft(i).extension
            × ft(i).dim(j).primary key
    End for
```

```
        to_delete = DENSITY(i) * |ft(i).extension|
        For j = 1 to to_delete do
            Random_delete(ft(i).extension)
        End for
        For j = 1 to |ft(i).extension| do
        // With |ft(i).extension| updated
            For k = 1 to NB_MEAS(i) do
                Ft(i).extension.tuple(j).measure(k)
                    = Random_float()
            End for
        End for
End for
```

Figure 5: DWEB fact tables generation algorithm

We use three main classes of functions and one procedure in these algorithms.

1. `Primary_key()`, `String_member()` and `Float_measure()` return attribute names for primary keys, members in hierarchy levels, and measures in fact tables, respectively. These names are labeled sequentially and prefixed by the table's name (e.g., DIM1_1_DESCR1, DIM1_1_DESCR2...).

2. `Integer_primary_key()`, `Random_key()`, `Random_string()` and `Random_float()` return sequential integers with respect to a given table (no duplicates are allowed), random instances of the specified table's primary key (random values for a foreign key), random strings of fixed size (20 characters) selected from a precomputed referential of strings and prefixed by the corresponding attribute name, and random single-precision real numbers, respectively.

3. `Random_dimension()` returns a dimension that is chosen among the existing dimensions that are not already describing the fact table in parameter.

4. `Random_delete()` deletes one tuple at random from the extension of a table.

Except in the `Random_delete()` procedure, where the random distribution is uniform, we use Gaussian random distributions to introduce a skew, so that some of the data, whether in the fact tables or the dimensions, are referenced more frequently than others as it is normally the case in real-life data warehouses.

Remark: The way density is managed in Figure 5 is grossly non-optimal. We chose to present the algorithm that way for the sake of clarity, but the actual implementation does not create all the tuples from the cartesian product, and then delete some of them. It directly generates the right number of tuples by using the density rate as a probability for each tuple to be created.

Workload Model

In a data warehouse benchmark, the workload may be subdivided into:

1. a load of decision-support queries (mostly OLAP queries);

2. the ETL (data generation and maintenance) process.

To design DWEB's workload, we inspire both from TPC-DS' workload definition (which is very elaborate) and information regarding data warehouse performance from other sources (BMC, 2000; Greenfield, 2004b). However, TPC-DS' workload is quite complex and somehow confusing. The reporting, ad-hoc decision-support and OLAP query classes are very similar, for instance, but none of them include any specific OLAP operator such as Cube or Rollup. Since we want to meet Gray's simplicity criterion, we propose a simpler workload. In particular, we do not address the issue of nested queries for now. Furthermore, we also have to design a workload that is consistent with the variable nature of the DWEB data warehouses.

We focus in this section on the definition of a query model that excludes update operations. The ETL and warehouse refreshing processes are addressed in the next section.

*Query Model*

The DWEB workload models two different classes of queries: purely decision-oriented queries involving common OLAP operations, such as cube, roll-up, drill-down and slice and dice; and extraction queries (simple join queries). We define our generic query model (Figure 6) as a grammar that is a subset of the SQL-99 standard, which introduces much-needed analytical capabilities to relational database querying. This increases the ability to perform dynamic, analytic SQL queries.

```
Query ::-
Select                  ![<Attribute Clause> | <Aggregate
                        Clause> | [<Attribute Clause>,
                        <Aggregate Clause>]]
From                    !<Table Clause> [<Where Clause>
                        || [<Group by Clause> * <Having
                        Clause>]]
Attribute Clause ::-    Attribute name [[, <Attribute
                        Clause>] | ⊥]
Aggregate Clause ::-    ![Aggregate function name
                        (Attribute name)] [As Alias] [[,
                        <Aggregate Clause>] | ⊥]
Table Clause ::-        Table name [[, <Table Clause>] |
                        ⊥]
Where Clause ::-        Where ![<Condition Clause> |
                        <Join Clause>| [<Condition
                        Clause> And <Join Clause>]
Condition Clause ::-    ![Attribute name <Comparison op-
                        erator> <Operand Clause>] [[<Log-
                        ical operator> <Condition
```

```
                            Clause>] | ⊥]
Operand Clause ::-          [Attribute name | Attribute value
                            | Attribute value list]
Join Clause ::-             ![Attribute name i = Attribute
                            name j] [[And <Join Clause>] | ⊥]
Group by Clause ::-         Group by [Cube | Rollup]
                            <Attribute Clause>
Having Clause ::-           [Alias | Aggregate function name
                            (Attribute name)] <Comparison
                            operator> [Attribute name |
                            Attribute value list]

Key:                        The [ and ] brackets are delimiters.
                            !<A>: A is required.
                            *<A>: A is optional.
                            <A ‖ B>: A or B.
                            <A | B>: A exclusive or B.
                            ⊥: empty clause.
                            SQL language elements are indicated in bold.
```

Figure 6: DWEB query model

*Parameterization*

DWEB's workload parameters help users tailor the benchmark's load, which is also dependent from the warehouse schema, to their needs. Just like DWEB's database parameter set (cf. previous section), DWEB's workload parameter set (Table 3) has been designed with Gray's simplicity criterion in mind. These parameters determine how the query model from Figure 6 is instantiated. These parameters help define workload size and complexity, by setting up the proportion of complex OLAP queries (i.e., the class of queries) in the workload , the number of aggregation operations, the presence of a Having clause in the query, or the number of subsequent drill-down operations.

| Parameter name | Meaning | Default value |
|---|---|---|
| *NB_Q* | Approximate number of queries in the workload | 100 |
| *AVG_NB_ATT* | Average number of selected attributes in a query | 5 |
| *AVG_NB_RESTR* | Average number of restrictions in a query | 3 |
| *PROB_OLAP* | Probability that the query type is OLAP | 0.9 |
| *PROB_EXTRACT* | Probability that the query is an extraction query | 1 - *PROB_OLAP* |
| *AVG_NB_AGGREG* | Average number of aggregations in an OLAP query | 3 |
| *PROB_CUBE* | Probability of an OLAP query to use the Cube operator | 0.3 |
| *PROB_ROLLUP* | Probability of an OLAP query to use the Rollup operator | 1 - *PROB_CUBE* |
| *PROB_HAVING* | Probability of an OLAP query to include a Having clause | 0.2 |
| *AVG_NB_DD* | Average number of drill-downs after an OLAP query | 3 |

Table 3: DWEB workload parameters

Here, we have only a limited number of high-level parameters (eight parameters, since *PROB_EXTRACT* and *PROB_ROLLUP* are derived from *PROB_OLAP* and *PROB_CUBE*, respectively). Indeed, it cannot be envisaged to dive further into detail if the workload is as large as several hundred queries, which is quite typical.

Remark: *NB_Q* is only an *approximate* number of queries because the number of drill-down operations after an OLAP query may vary. Hence we can stop generating queries only when we actually have generated as many or more queries than *NB_Q*.

*Generation Algorithm*

The pseudo-code of DWEB's workload generation algorithm is presented in Figures 7a and 7b. The algorithm's purpose is to generate a set of SQL-99 queries that can be directly executed on the synthetic data warehouse defined in the previous section. It is subdivided into two steps:

1. generate an initial query that may be either an OLAP or an extraction (join) query;

2. if the initial query is an OLAP query, execute a certain number of drill-down operations based on the first OLAP query. More precisely, each time a drill-down is performed, a member from a lower level of dimension hierarchy is added to the attribute clause of the previous query.

Step 1 is further subdivided into three substeps:

1. the Select, From, and Where clauses of a query are generated simultaneously by randomly selecting a fact table and dimensions, including a hierarchy level within a given dimension hierarchy;

2. the Where clause is supplemented with additional conditions;

3. eventually, it is decided whether the query is an OLAP query or an extraction query. In the second case, the query is complete. In the first case, aggregate functions applied to measures of the fact table are added in the query, as well as a Group by clause that may include either the Cube or the Rollup operator. A Having clause may optionally be added in too. The aggregate function we apply on measures is always Sum since it is the most common aggregate in cubes. Furthermore, other aggregate functions bear similar time complexities, so they would not bring in any more insight in a performance study.

```
n = 0
While n < NB_Q do
    // Step 1: Initial query
    // Step 1.2: Select, From and Where clauses
    i = Random_FT() // Fact table selection
    attribute_list = Ø
    table_list = ft(i)
    condition_list = Ø
    For k = 1 to Random_int(AVG_NB_ATT) do
        // Dimension selection
        j = Random_dimension(ft(i))
        l = Random_int(1, ft(i).dim(j).nb_levels)
        // Positioning on hierarchy level l
        hl = ft(i).dim(j) // Current hierarchy level
        m = 1 // Level counter
        fk = ft(i).intention.primary_key.element(j)
        // This foreign key corresponds to
        // ft(i).dim(j).primary_key
        While m < l and hl.child ≠ NIL do
            // Build join
            table_list = table_list ∪ hl
            condition_list = condition_list
                ∪ (fk = hl.intention.primary_key)
            // Next level
            fk = hl.intention.foreign_key
            m = m + 1
            hl = hl.child
        End while
        attribute_list = attribute_list
            ∪ Random_attribute(hl.intention)
    End for
    // Step 1.2: Supplement Where clause
    For k = 1 to Random_int(AVG_NB_RESTR) do
        condition_list = condition_list
            ∪ (Random_attribute(attribute_list)
                = Random_string())
    End for
    // Step 1.3: OLAP or extraction query selection
    p1 = Random_float(0, 1)
    If p1 ≤ PROB_OLAP then // OLAP query
        // Aggregate clause
        aggregate_list = Ø
        For k = 1 to Random_int(AVG_NB_AGGREG) do
            aggregate_list = aggregate_list
                ∪ (Random_measure(ft(i).intention)
        End for
                                                    ../..
```

Figure 7a: DWEB workload generation algorithm

```
../..
        // Group by clause
        group_by_list = attribute_list
        p2 = Random_float(0, 1)
        If p2 ≤ PROB_CUBE then
            group_by_operator = CUBE
        Else
            group_by_operator = ROLLUP
        End if
```

```
            // Having clause
            P3 = Random_float(0, 1)
            If p3 ≤ PROB_HAVING then
                having_clause
                    = (Random_attribute(aggregate_list), ≥,
                        Random_float())
            Else
                having_clause = ∅
            End if
        Else // Extraction query
            group_by_list = ∅
            group_by_operator = ∅
            having_clause = ∅
        End if
        // SQL query generation
        Gen_query(attribute_list, aggregate_list, table_list,
            condition_list, group_by_list, group_by_operator,
            having_clause)
        n = n + 1
        // Step 2: Possible subsequent DRILL-DOWN queries
        If p1 ≤ PROB_OLAP then
            k = 0
            While k < Random_int(AVG_NB_DD)
                and hl.parent ≠ NIL do
                k = k + 1
                hl = hl.parent
                att = Random_attribute(hl.intention)
                attribute_list = attribute_list ∪ att
                group_by_list = group_by_list ∪ att
                Gen_query(attribute_list, aggregate_list,
                table_list, condition_list, group_by_list,
                group_by_operator, having_clause)
            End while
            n = n + k
        End if
End while
```

Figure 7b: DWEB workload generation algorithm (continued)

We use three classes of functions and a procedure in this algorithm.

1. `Random_string()` and `Random_float()` are the same functions than those already described in the Database Generation section. However, we introduce the possibility for `Random_float()` to use either a uniform or a Gaussian random distribution. This depends on the function parameters: either a range of values (uniform) or an average value (Gaussian). Finally, we introduce the `Random_int()` function that behaves just like `Random_float()` but returns integer values.

2. `Random_FT()` and `Random_dimension()` help select a fact table or a dimension describing a given fact table, respectively. They both use a Gaussian random distribution, which introduces an access skew at the fact table and dimension levels. `Random_dimension()` is also already described in the Database Generation section.

3. `Random_attribute()` and `Random_measure()` are very close in behavior. They return an attribute or a measure, respectively, from a table intention or a list of attributes. They both use a Gaussian random distribution.

4. `Gen_query()` is the procedure that actually generates the SQL-99 code of the workload queries, given all the parameters that are needed to instantiate our query model.

ETL Process

When designing DWEB's ETL process, we have to consider again the relevance *vs.* simplicity tradeoff (cf. Gray's criteria). Though the ETL phase may take up to 80% of the time devoted to a data warehousing project, it would not be reasonable to include its full complexity in a benchmark tool. Hence, we balanced in favor of simplicity. However, we present here a first step toward including a full ETL phase into DWEB; extensions are definitely possible.

*Model*

Since the DWEB software is a standalone tool that generates data and workloads, we chose not to include an extraction phase in its ETL capability. Data updates are performed directly in the database to keep DWEB's usage simple and minimize external file management. However, data updates could also easily be recorded into flat files before being applied, to simulate an extraction phase.

We did not include any transformation in the process either, despite it is a very important phase in the ETL process. However, in a benchmark, such transformations are simulated to consume CPU time (this is the tactic adopted in TPC-DS). In DWEB, we consider that the processing time of various tests in the insert and modify procedures related to the loading phase might be considered as equivalent to simulating transformations.

We thus focus on the loading phase. A data warehouse schema is built on two different concepts: facts and dimensions. Updates might be insertions, modifications or deletions. Since data are normally historicized in a data warehouse, we do not consider deletions. Hence, we can identify four warehouse refreshing types for which we adopt specific strategies.

1. *Insertions in fact tables* are simple. They involve few constraints at the schema level, save that we cannot use an existing primary key in the related fact table. To achieve an insertion, we randomly fetch one primary key in each dimension to build an aggregate fact table key, and then add random measure values to complete the fact.

2. *Insertions in dimensions* raise a new issue. They must be dispatched in all hierarchy levels. Hence, for each dimension, we seek to insert elements from the highest hierarchy level (coarsest granularity grain) into the lowest hierarchy level (finest granularity grain). New dimension members only need a new, unique primary key to be inserted in a given hierarchy level.

3. *Modifications in fact tables* only necessitate randomly fetching an existing fact and modifying its measure values.

4. *Modifications in dimensions* must finally take dimension hierarchy levels in into account to avoid introducing inconsistencies in the hierarchy.

*Parameterization*

DWEB's ETL parameters help users tune how the data warehouse is refreshed. Like the other parameters in DWEB, they have been designed with Gray's simplicity criterion in mind. These parameters direct how the ETL model is applied. They basically define refresh and insertion/modification rates. We voluntarily define only a small number (three, since *FRR* and *MR* are derived from *DRR* and *IR*, respectively) of high-level parameters (Table 4).

| Parameter name | Meaning | Default value |
|---|---|---|
| *GRR* | Global refresh rate | 0.01 |
| *DRR* | Dimension refresh rate | 0.05 |
| *FRR* | Fact refresh rate | 1 – DRR |
| *IR* | Insert rate | 0.95 |
| *MR* | Modification rate | 1 – IR |

Table 4: DWEB ETL parameters

*GRR* represents the total number of records from fact and dimension tables that must be refreshed (insertion and modifications included), with respect to current warehouse size. *DRR* and *FRR* control the proportion of these updates that are performed on dimension and fact tables, respectively. Finally, *IR* and *MR* control the proportion of insertions and modifications, respectively.

*Refresh Algorithms*

The refresh phase in <mark>DWEB</mark> is actually achieved with the help of two refresh procedures, one for dimensions and one for fact tables. Their pseudo-code is presented in Figures 8 and 9, respectively. Both procedures follow the same principle:

1. compute the number of tuples to insert or update with respect to parameters *GRR*, *DRR*, *FRR*, *IR*, and *MR*, as well as the total number of tuples in the warehouse *GLOBAL_SIZE*;

2. insert or modify as many tuples in the corresponding table — modifications affect a randomly selected tuple. Furthermore, dimension updates are dispatched on all hierarchy levels.

```
For i = 1 to TOT_NB_DIM do
    ins_nb = ((GLOBAL_SIZE * GRR * DRR * IR)
        / TOT_NB_DIM) / NB_LEVELS(i)
    mod_nb = ((GLOBAL_SIZE * GRR * DRR * MR)
        / TOT_NB_DIM) / NB_LEVELS(i)
    For j = NB_LEVELS(i) to 1 step -1 do
        // Insertions
        For k = 1 to ins_nb do
            Insert_into_Dim(dim(i).level(j))
        End for
        // Modifications
        For k = 1 to mod_nb do
            Modify_Dim(Random_Key(dim(i).level(j))
        End for
    End for
End for
```

Figure 8: DWEB dimension refresh procedure

```
For i = 1 to NB_FT do
    For j = 1 to (GLOBAL_SIZE * GRR * FRR * IR)
        / |ft(i).extension| do
        Insert_into_FT(ft(i))
    End for
    For j = 1 to (GLOBAL_SIZE * GRR * FRR * MR)
        / |ft(i).extension| do
        Modify_FT(Random_Key(ft(i))
    End for
End for
```

Figure 9: DWEB fact table refresh procedure

We use two new classes of procedures in this algorithm.

1. `Insert_into_Dim()` and `Insert_into_FT()` insert new tuples into dimension and fact tables, respectively. The main difference between these two procedures is that dimension insertion manages foreign key selection for pointing to the next hierarchy level, whereas there is no such constraint in a fact table.

2. `Modify_Dim()` and `Modify_FT()` modify one tuple, identified by its primary key, from a dimension or fact table, respectively. Primary keys are provided by the `Random_key()` function (cf. Database Model section), which returns random instances of the specified table's primary key. `Modify_Dim()` and `Modify_FT()` only differ by the updated attribute's nature:

dimension members are strings, while fact measures are numerical. They are generated with the `Random_string()` and `Random_float()` functions, respectively (cf. Database Model section).

Execution Protocol

*Protocol*

DWEB's test protocol is quite classical for a benchmark, and is actually a variation of TPC-DS'. It is constituted of two distinct parts:

1. a *load test* that consists in filling the data warehouse structure with data;

2. a *performance test* that evaluates system response and that is further subdivided into two steps:

2.1. a *cold run* in which the workload is applied onto the test warehouse once;

2.2. a *warm run* that is replicated *REPN* times and that includes the warehouse refresh process and another execution of the workload.

The pseudo-code for the performance test is presented in Figure 10. The main difference between DWEB's and TPC-DS' execution protocols is that DWEB's warm run may be executed many times instead of just one.

```
// Cold run
etime[0] = time()
Execute_Workload()
etime[0] = time() - etime[0]
// Warm run
For i = 1 to REPN do
    rtime[i] = time()
    Execute_Refresh(GLOBAL_SIZE)
    rtime[i] = time() - rtime[i]
    etime[i] = time()
    Execute_Workload()
    etime[i] = time() - etime[i]
End for
```
Figure 10: Performance test algorithm

Remark: The *GRR* parameter may be set to zero if users do not want to include warehouse refresh tests.

*Performance Metric*

The performance metric we retained in DWEB is *response time*. It is computed separately for workload execution and data warehouse refreshing, so that any run time (e.g., cold run time, refresh time in warm run replication #*i*…) can be displayed. Global, average, minimum and maximum execution times are also computed, as well standard deviation. Note that this kind of atomic approach for assessing performance allows to derive any more complex, composite metrics, such as TPC-H's and TPC-D's, if necessary.

## DWEB TUTORIAL

We present in this section a short tutorial that illustrates DWEB's usage and shows how DWEB's execution protocol is implemented in practice. DWEB is a Java application. Its main GUI (Graphical User Interface) is depicted in Figure 11. It is divided into three sections/panels:

1. *database connection:* JDBC (Java Database Connectivity) connection to a database server and database;

2. *action:* the actual benchmark interface that helps set parameters and launch tests;

3. *information:* this window displays messages when an event or error occurs.

Actually using DWEB through the "Action" panel is a four-step process.

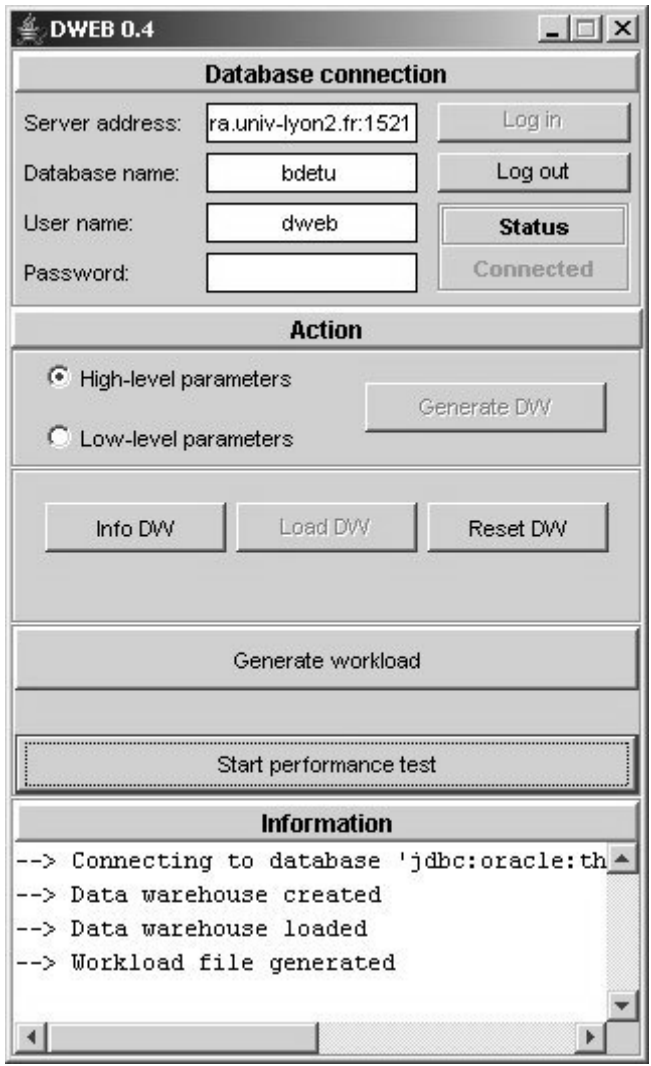

Figure 11: DWEB GUI

Data Warehouse Generation

Clicking on the "Generate DW" command button helps set either the full range of low-level parameters or only the high-level parameters (Figure 12), which we recommend for most performance tests. Then, the data warehouse's (empty) structure is automatically created.

Figure 12: DWEB database parameterization

Load Test

The second subpanel in the "Action" panel features three command buttons. Since DWEB's parameters might sound abstract, we provide through the "Info DW" command button an estimation of data warehouse size in megabytes before it is actually loaded. Hence, users can reset the parameters to better represent the kind of warehouse they need, if necessary.

The "Load DW" command button actually launches the load test, whose status is displayed to the user (Figure 13), who can interrupt the process at any time. When the data warehouse is loaded, load time is displayed.

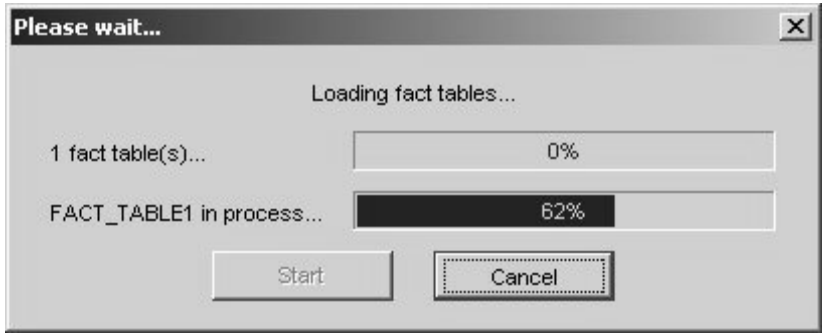

Figure 13: DWEB load test in process

Finally, the "Reset DW" command button helps destroy the current data warehouse. Since table names are standard in DWEB, this feature helps avoiding name conflicts when generating a new data warehouse. If several warehouses need to be stored concurrently, several different database users must be created for this sake.

Workload Generation

Workload generation is simply achieved by clicking on the "Generate workload" command button, which triggers workload parameter setup (Figure 14) and save its queries into an external file, so that they can be reused.

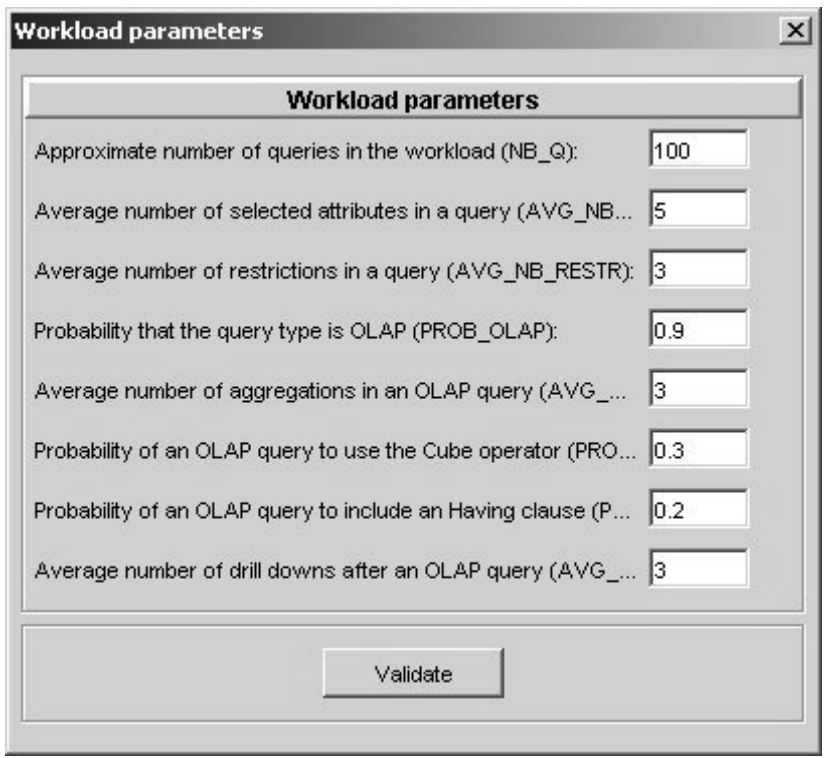

Figure 14: DWEB workload parameterization

Performance Test

Finally, the "Start performance test" command button helps set the new ETL and protocol parameters (cf. previous section). They are then recapitulated in the performance test window (Figure 15) that actually allows launching benchmark execution. Every workload execution and refresh operation time is displayed and also recorded separately in a CSV (Comma-Separated Values) file that can later be processed in a spreadsheet or any other application. Warm run total, average, minimum and maximum times, as well as standard deviation, for refresh operations, workload executions, and both (refresh + workload total), are computed. Performance tests may be reiterated any number of times, with or without generating a new workload each time.

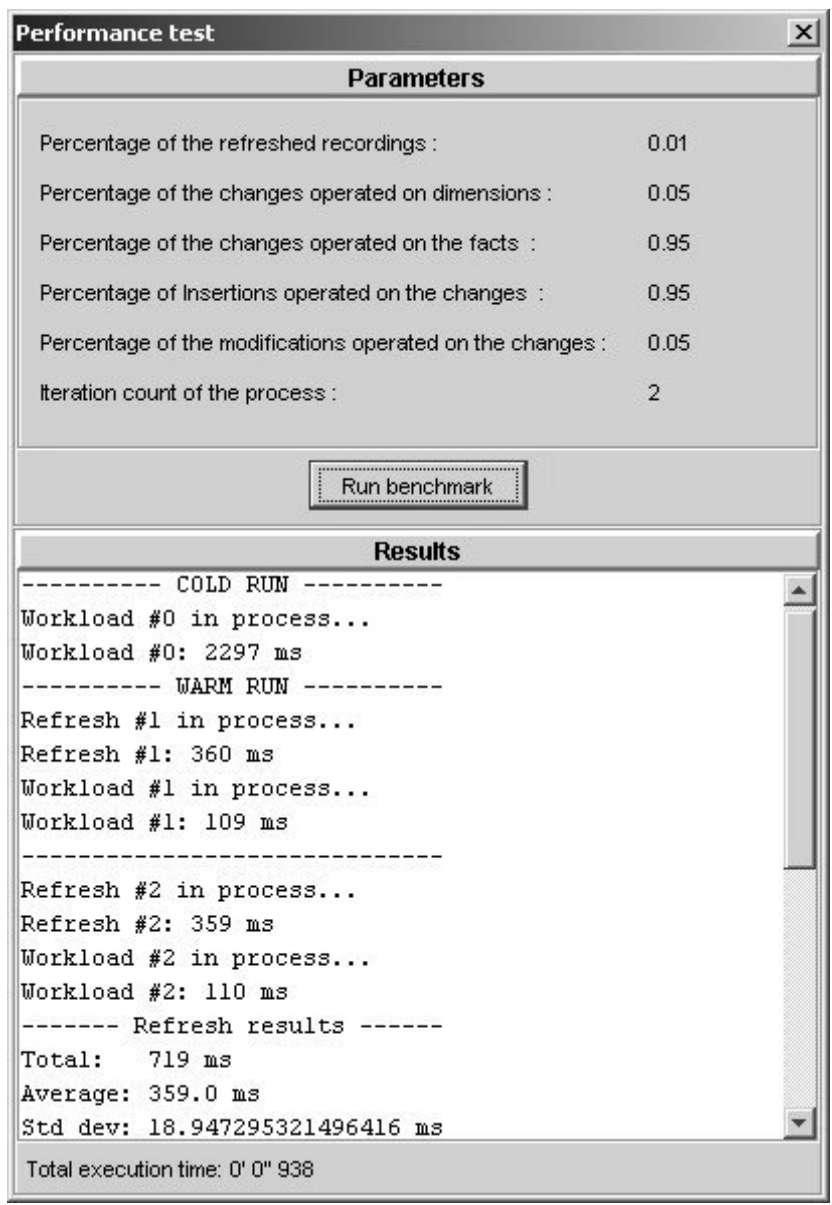

Figure 15: DWEB performance test window

## CONCLUSION AND PERSPECTIVES

We have mainly proposed DWEB, which is currently the only operational data warehouse benchmark to the best of our knowledge, to help data warehouse designers select among alternate architectures and/or performance optimization techniques. However, it can also be used, as the TPC benchmarks, for sheer performance comparisons. It is indeed possible to save a given warehouse and its associated workload to run tests on different systems and/or with various optimization techniques.

To satisfy the relevance and adaptability criteria, DWEB can generate various ad-hoc synthetic data warehouses and their associated workloads. Popular data warehouse schemas, such as star, snowflake, and constellation schemas, as well as much-used decision-support operators such as cube, roll-up or drill-down, are indeed supported by our tool. These features are piloted by a full set of low-level parameters, but we have also proposed a series of high-level parameters that are limited in number, not to sacrifice too much Gray's simplicity criterion. Finally, we have opted to implement DWEB with the Java language to satisfy the portability criterion. DWEB's code is freely available on-line (3).

We have illustrated sample usages of DWEB by evaluating the efficiency of several indexing techniques on various data warehouse configurations (Darmont *et al.*, 2005; Darmont *et al.*, 2007). Though these experiments were not actually new, they helped us demonstrate DWEB's relevance. We indeed obtained results that were consistent with previously published results regarding bitmap join indices (O'Neil & Graefe, 1995) and star-join indices (Bellatreche *et al.*, 2002). We could underline again the crucial nature of indexing choices in data warehouses. Furthermore, since such choices depend on the warehouse's architecture, we showed DWEB's usefulness in a context where "mono-schema" benchmarks are not relevant.

Our work opens up many perspectives for further developing and enhancing DWEB. First, the warehouse metamodel and query model are currently deliberately simple. They could definitely be extended to be more representative of real data warehouses, i.e., more relevant. For example, the warehouse metamodel could feature many-to-many relationships between dimensions and fact tables, and hierarchy levels that are shared by several dimensions. Our query model could also be extended with more complex queries such as nested queries that are common in OLAP usages. Similarly, our DWEB's ETL feature focuses on the loading phase, and could be complemented by extraction and transformation capabilities. TPC-DS' specifications and other existing studies (Labrinidis & Roussopoulos, 1998) could help us complementing our own tool.

We have also proposed a set of parameters for DWEB that suit both the models we developed and our expected usage of the benchmark. However, a formal validation would help select the soundest parameters. More experiments should also help us to evaluate the pertinence of our parameters and maybe propose sounder default values. Other parameters could also be considered, such as the domain cardinality of hierarchy attributes or the selectivity factors of restriction predicates in queries. This kind of information may indeed help designers to choose an architecture that supports some optimization techniques adequately.

Finally, we only used response time as a performance metric. Other metrics must be envisaged, such as the metrics designed to measure the quality of data warehouse conceptual models (Serrano *et al.*, 2003; Serrano *et al.*, 2004). Formally validating these metrics would also improve DWEB's usefulness.

ACKNOWLEDGEMENTS

The author would like to thank Audrey Gutierrez and Benjamin Variot for coding DWEB's latest improvements.

ENDNOTES

(1) http://www.tpc.org

(2) http://cgmlab.cs.dal.ca/downloadarea/datasets/

(3) http://bdd.univ-lyon2.fr/download/dweb.tgz

REFERENCES

Ballinger, C. (1993). TCP-D: Benchmarking for Decision Support. *The Benchmark Handbook for Database and Transaction Processing Systems*, second edition. Morgan Kaufmann.

Bellatreche, L., Karlapalem, K., & Mohania, M. (2002). Some issues in design of data warehousing systems. *Data warehousing and web engineering*. IRM Press. 22-76.

Bhashyam, R. (1996). TCP-D: The Challenges, Issues and Results. *22th International Conference on Very Large Data Bases*, Mumbai (Bombay), India. *SIGMOD Record*. 4, 593.

BMC Software. (2000). Performance Management of a Data Warehouse. *http://www.bmc.com*

Darmont, J., Bentayeb, F., & Boussaïd, O. (2005). DWEB: A Data Warehouse Engineering Benchmark. *7th International Conference on Data Warehousing and Knowledge Discovery (DaWaK 05)*, Copenhagen, Denmark. *LNCS*. 3589, 85-94.

Darmont, J., Bentayeb, F., & Boussaïd, O. (2007). Benchmarking Data Warehouses. *International Journal of Business Intelligence and Data Mining*. 2(1), 79-104.

Darmont, J., & Schneider, M. (2000). Benchmarking OODBs with a Generic Tool. *Journal of Database Management*. 11(3), 16-27.

Demarest, M. (1995). A Data Warehouse Evaluation Model. *Oracle Technical Journal*. 1(1), 29.
Gray, J., Ed. (1993). *The Benchmark Handbook for Database and Transaction Processing Systems,* second edition. Morgan Kaufmann.

Greenfield, L. (2004). Performing Data Warehouse Software Evaluations.
*http://www.dwinfocenter.org/evals.html*

Greenfield, L. (2004). What to Learn About in Order to Speed Up Data Warehouse Querying.
*http://www.dwinfocenter.org/fstquery.html*

Inmon, W.H. (2002). *Building the Data Warehouse*, third edition. John Wiley & Sons.

Kimball, R., & Ross, M. (2002). *The Data Warehouse Toolkit: The Complete Guide to Dimensional Modeling*, second edition. John Wiley & Sons.

Labrinidis, A., & Roussopoulos, N. (1998). *A Performance Evaluation of Online Warehouse Update Algorithms*. Technical report CS-TR-3954. Department of Computer Science, University of Maryland.

OMG. (2003). *Common Warehouse Metamodel (CWM) Specification version 1.1*. Object Management Group.

O'Neil, P.E., & Graefe, G. (1995). Multi-table joins through bitmapped join indices. *SIGMOD Record*. 24(3), 8-11.

Pendse, N. (2003). The OLAP Report: How not to buy an OLAP product.
*http://www.olapreport.com/How_not_to_buy.htm*

Poess, M., & Floyd, C. (2000). New TPC Benchmarks for Decision Support and Web Commerce. *SIGMOD Record*. 29(4), 64-71.

Poess, M., Smith, B., Kollar, L., & Larson, P.A. (2002). TPC-DS: Taking Decision Support Benchmarking to the Next Level. *2002 ACM SIGMOD International Conference on Management of Data*, Madison, Wisconsin, USA. 582-587.

Poole, J., Chang, D., Tolbert, D., & Mellor, D. (2003). *Common Warehouse Metamodel Developer's Guide*. John Wiley & Sons.

Serrano, M., Calero, C., & Piattini, M. (2003). Metrics for Data Warehouse Quality. *Effective Databases for Text & Document Management*. 156-173.

Serrano, M., Calero, C., Trujillo, J., Lujan-Mora, S., & Piattini, M. (2004). Towards a Metrics Suite for Conceptual Models of Datawarehouses. *1st International Workshop on Software Audit and Metrics (SAM 04)*, Porto, Portugal. 105-117.

Thomsen, E. (1998). Comparing different approaches to OLAP calculations as revealed in benchmarks. *Intelligence Enterprise's Database Programming & Design*.
http://www.dbpd.com/vault/9805desc.htm

TPC. (1998). *TPC Benchmark D Standard Specification Version 2.1*. Transaction Processing Performance Council.

TPC. (2003). *TPC Benchmark R Standard Specification Revision 2.1.0*. Transaction Processing Performance Council.

TPC. (2006). *TPC Benchmark H Standard Specification Revision 2.6.1*. Transaction Processing Performance Council.

TPC. (2007). *TPC Benchmark DS Standard Specification Draft version 52*. Transaction Processing Performance Council.